# Sources of false positives and false negatives in the STATCHECK algorithm: Reply to Nuijten et al. (2016)

arXiv:1610.01010v8 [q-bio.NC]

Thomas Schmidt
University of Kaiserslautern, Germany
Experimental Psychology Unit
www.sowi.uni-kl.de/psychologie
thomas.schmidt@sowi.uni-kl.de

## Abstract

**STATCHECK is an *R* algorithm designed to scan papers automatically for inconsistencies between test statistics and their associated $p$ values (Nuijten et al., 2016). The goal of this comment is to point out an important and well-documented flaw in this busily applied algorithm: It cannot handle corrected $p$ values. As a result, statistical tests applying appropriate corrections to the $p$ value (e.g., for multiple tests, post-hoc tests, violations of assumptions, etc.) are likely to be flagged as reporting inconsistent statistics, whereas papers omitting necessary corrections are certified as correct. The STATCHECK algorithm is thus valid for only a subset of scientific papers, and conclusions about the quality or integrity of statistical reports should never be based solely on this program.**

STATCHECK is an *R* algorithm designed to scan papers automatically for inconsistencies between test statistics and their reported $p$ values (Nuijten, Hartgerink, van Assen, Epskamp, and Wicherts, 2016). The algorithm is currently part of these authors' project to automatically scan thousands of scientific psychology papers and post the results on a website when the program detects some discrepancy, essentially flagging papers for reporting inconsistent test statistics.

This project is undertaken by researchers who undoubtedly have the best intentions and whose work requires no small amount of personal courage. I do object about the fact that the comments are made anonymously, are obviously issued without the papers being read, are not corrected when turning out to be false, and are issued without reliably notifying the authors, which is all ethically problematic. I also believe it is unwise to place the commentaries on the PubPeer website (www.pubpeer.com), which has heretofore been known for revealing data forgery and therefore operates in strict anonymity. This leaves authors flagged by the STATCHECK project (as well as the general public) unclear about whether they are receiving a friendly reminder for more careful copytyping or whether they are being accused of scientific misconduct.

However, the ethical and political implications of the STATCHECK project should be considered apart from the actual properties of the program. If the program is employed for seeking out papers that report their statistics unreliably, the question arises whether STATCHECK reports are accurate enough for such a purpose, especially when the results are made public. In this commentary, I want to focus on a single problem that renders the STATCHECK algorithm unsuitable for this purpose: It cannot handle corrected *p* values. Therefore, it can be validly applied only to a subset of the psychological research literature.

What kind of corrections are we talking about? Statistical tests rarely appear in isolation, but typically in the context of other tests. Also, they usually make assumptions about the underlying stochastical structure of the data, and those assumptions are often violated to some degree. Therefore, it is often necessary to correct the results for multiple testing (e.g., Bonferroni corrections, Tukey tests), for making tests post-hoc instead of planning them beforehand (e.g., Scheffé tests), and for violations of stochastical assumptions (e.g., Greenhouse-Geisser or Huynh-Feldt corrections).[1]

To understand why this leads to problems for STATCHECK, consider the well-known Greenhouse-Geisser correction for a repeated-measures *F* test in analysis of variance. This correction is designed to make the test more conservative when the data violate an esoteric assumption called sphericity (Box, 1954; Geisser & Greenhouse, 1958; I prefer the more precise correction proposed by Huynh & Feldt, 1970). This is accomplished by multiplying the degrees of freedom (*df*) for the *F* test by a correction term named $\varepsilon$, which is a number between zero and one. This leaves the *F* value unchanged, but it increases the resulting *p* value and turns the degrees of freedom from integers to fractional numbers. Because fractional degrees of freedom are confusing and make it difficult to see which exact hypothesis is tested, it is customary to report the (unaffected) *F* statistic, the *original* degrees of freedom, and the *corrected p* value, and state that the Greenhouse-Geisser correction was applied.[2]

How does STATCHECK deal with this? In fact, the algorithm is simply a string detection program with a few statistical tables (Nuijten et al., 2016, pp. 1206-1207). It looks for a string that reads "*test statistic (df)* = x, *p* = y", taking into account some variations in typography. Then it recalculates the *p* value from the test statistic and the degrees of freedom and compares it with the *p* value in the manuscript. In the case of our Greenhouse-Geisser correction, it will detect an inconsistency because the *F* value, looked up in the table with the uncorrected degrees of freedom, leads to a smaller *p* value than the one we stated. In case the paper is selected for public scrutiny by the STATCHECK project, the program output is posted on the PubPeer website, and an attempt is made to send an e-mail to the authors that their paper has received a comment.[3] In the case of our recent paper (Schmidt, Hauch, & Schmidt, 2015) that mail reads like this:

> "Using the R package statcheck (v1.0.1), the HTML version of this article was scanned on 2016-08-05 for statistical results (t, r, F, Chi2, and Z values) reported in APA format (for specifics, see Nuijten et al., 2016). An automatically generated report follows.
> The scan detected 86 statistical results in APA format, of which 18 contained potentially incorrect statistical results, of which 2 may change statistical significance (alpha = .05). Potential one-tailed results were taken into account when 'one-sided', 'one-tailed', or 'directional' occurred in the text.

---

[1] This is just a sample of correction problems, mostly concerning analysis of variance. The use of statistical correction in correlational studies is much more complex.

[2] An anonymous comment on PubPeer that $\varepsilon$-corrected *F* tests generally have to be reported with fractional degrees of freedom is incorrect: Even though this is occasionally seen and is certainly not wrong, it has been uncommon practice in the timespan covered by Nuijten et al.'s (2016) paper.

[3] It seems that this last step fails when an author changes affiliation. It is therefore advisable for everybody to check PubPeer regularly for new entries.

The errors that may change statistical significance were reported as:
F (3, 21) = 3.37, p = .053 (recalculated p-value: 0.03775)
F (1, 7) = 3.74, p = .035 (recalculated p-value: 0.09438)

The errors that may affect the computed p-value (but not the statistical significance) were reported as:
F (3, 21) = 6.47, p = .017 (recalculated p-value: 0.00284)
F (3, 21) = 4.01, p = .028 (recalculated p-value: 0.02106)
F (3, 21) = 6.89, p = .018 (recalculated p-value: 0.00208)
F (3, 21) = 4.39, p = .037 (recalculated p-value: 0.01509)
F (2, 14) = 14.09, p = .007 (recalculated p-value: 0.00044)
F (6, 42) = 8.81, p = .021 (recalculated p-value: 0)
F (3, 21) = 6.07, p = .002 (recalculated p-value: 0.00384)
F (3, 21) = 9.54, p = .002 (recalculated p-value: 0.00036)
F (3, 21) = 8.95, p = .012 (recalculated p-value: 0.00051)
F (3, 21) = 5.38, p = .020 (recalculated p-value: 0.00661)
F (3, 21) = 3.32, p = .041 (recalculated p-value: 0.03956)
F (1, 7) = 12.19, p < .001 (recalculated p-value: 0.01011)
F (1, 7) = 8.77, p = .003 (recalculated p-value: 0.02106)
F (3, 21) = 7.56, p = .003 (recalculated p-value: 0.0013)
F (3, 21) = 12.50, p = .002 (recalculated p-value: 7e-05)
2 (1) = 0.13, p = .929 (recalculated p-value: 0.71843)

Note that these are not definitive results and require manual inspection to definitively assess whether results are erroneous."

Apart from the last sentence, this report is inaccurate and should be read more carefully than STATCHECK read our paper. First, note that the incidence of mismatching *p* values far exceeds the number of honest mistakes expected from the most untalented of typewriters. Second, the incriminated *p* values tend to be substantially larger than the ones recalculated by the program. This is because we report all *p* values with a Huynh-Feldt correction but give the original degrees of freedom. Third, there is an exotic test statistic named "2", which has one degree of freedom and whose *p* value is recalculated from .9 to .7 by a method that would seriously interest me. (Note that this misread of a chi-squared statistic would be sufficient to flag the paper.) Fourth, one *p* value is being recalculated from .021 to exactly zero. It would be easy to ridicule all this, but no fewer than 4 out of the 98 *p* values we report are actually smaller than calculated by the program. It turns out that those four are real mistakes on our part (in one case, the *p* value should be .004 instead of .002, one case gives the wrong *F* value, and two cases give the wrong degrees of freedom), showing that the algorithm would be able to weed out some errors when applied during manuscript preparation. The problem here, however, is that the number of false alarms far exceeds the number of hits. In this report, STATCHECK detected 86 of our 98 significance tests, scoring 4 hits and 14 false alarms. In the second report we received on PubPeer (Seydell-Greenwald & Schmidt, 2012), the program performed substantially worse, detecting only 51 of our 82 tests and scoring 1 hit and 16 false alarms. Of the 5 true mistakes, 2 reflected errors in *p* values that might have affected statistical significance but fortunately did not, 2 were errors in degrees of freedom, and 1 was a wrong *F* value. Those are five typos out of 720 numbers reported.

The automated message is followed by a reference to the preprint of Nuijten et al. (2016), which was the only hint leading me to the persons responsible for the anonymous posting. I e-mailed Michèle Nuijten, a co-author of the algorithm, that our paper had been flagged for reporting *F* tests with uncorrected degrees of freedom but with Hunyh-Feldt corrected *p* values. Michèle responded friendly to my angry and sarcastic mail, apologizing for the false alarms but confirming that "statcheck can indeed not handle corrected *p*-values" (M. B. Nuijten, personal e-mail, September 28, 2016). This is also acknowledged in the original paper, which states that "statcheck does not take into account *p*-values that are adjusted for multiple testing" (Nuijten et al., 2016, p. 1206; see also their Appendix A for problems with Huynh-Feldt corrections). However, the authors do not view this as a problem because "when we automatically searched our sample of 37,717 articles, we found that only 96 articles reported the string 'Bonferroni' (0.3 %) and nine articles reported the string 'Huynh-Feldt' or 'Huynh Feldt' (0.03 %)" (Nuijten et al., 2016, p. 1207). But are these estimates plausible? In my own PDF folder of about 800 papers, which certainly differs systematically from their sample, 7.6 % of papers contain the word "Bonferroni". The string "post-hoc" (13.7 %) also appears quite frequently. The search strings "Greenhouse-Geisser" (1.5 %) and "Huynh-Feldt" (0.8 %) are indeed alarmingly rare, but both "Bonferroni" and "Huynh-Feldt" are about 25 times more frequent in my ad-hoc collection than in Nuijten et al.'s sample.

Interestingly, the numbers also clash with their own "validation check" (Nuijten et al., 2016, Appendix A) where the authors run the STATCHECK program on the sample of 49 of papers investigated by hand by Wicherts, Bakker, & Molenaar (2011). Surprisingly, that small sample happens to contain two papers with Huynh-Feldt-corrected tests. If the prevalence of Huynh-Feldt corrections were indeed 0.03 % as the authors suggest, then the binomial probability of finding two or more of these in a sample of 49 is only .000105, or one in 10,000, strongly suggesting some kind or miscount.[4] In that case, testing procedures problematic for STATCHECK would be more frequent than the authors suggest, and the program would produce more *false alarms* than they assume. At the same time, it is clear that those correction procedures are still vastly underused, and many of those uncorrected *p* values will be misleading. Those mistakes will all be *missed* by the STATCHECK algorithm. Therefore, the authors state correctly that "statcheck will miss some reported results and will incorrectly earmark some correct *p*-values as a reporting error" (Nuijten et al., 2016, p. 1207). This clear message does not seem to have reached some enthusiastic Twitter followers (hashtag: #*statcheck*) who brag about their "health certificate" because their studies had passed the STATCHECK test.

In fact, it is possible that they should be worried by that result. Corrections to the *p* value are common, and it is actually difficult to imagine a paper where such a correction is never needed. For instance, corrected *p* values should appear in any paper that reports analysis of variance on repeated measures, unless (1) all the factors involved have only two levels or (2) the sphericity assumption is met for each main effect and interaction. Whenever these special conditions do not apply, the *F* and *p* statistics would be expected to *mismatch* when the correction is done correctly, and to *match* if it is neglected (unless the *df* are presented as decimals). Similarly, many procedures correcting *p* values for multiple or post-hoc tests should lead to mismatch when applied *correctly*, whereas a match may indicate the *lack* of such a correction (all of this depending on the exact procedure and program package used). This would be the exact opposite of what the program seems to indicate.

---

[4] Fortunately, it is not at all necessary to scan all the 37,000 papers to assess the authors' claims. The *negative binomial distribution* gives the probability that $n + k - 1$ papers have to be randomly selected before the occurrence of the *k*th one that employs Bonferroni correction, given the null hypothesis that the true incidence is $p = 96/37,717$. Specifically, $P(K = k) = (k + n - 1 \text{ choose } k) \cdot p^k \cdot (1 - p)^n$. This method allows for a self-terminating search.

I would like to draw the following conclusions. 1) The STATCHECK algorithm will be useful during manuscript production to seek out additional typing and copying errors, provided that the manuscript has already been carefully checked. 2) It has to be applied with the understanding that the output contains misses as well as false alarms, and likely more than suggested in the paper. 3) The algorithm occasionally misreads or misinterprets statistics, and papers may be flagged on PubPeer just because of that. 4) Statistical tests applying appropriate corrections to the *p* value are likely to be flagged as reporting inconsistent statistics, whereas papers omitting all necessary corrections are likely to be certified as correct.

In sum, the algorithm is only valid for a subset of scientific papers. It is unsuitable for the automated scanning of large numbers of papers, without any human being carefully reading the method sections and understanding the logical and statistical context in which statistical numbers are being applied. Being flagged by the program does not imply that a single mistake was committed, let alone any form of scientific misconduct; and being cleared by the program does not imply that the statistics are sound. Conclusions about the quality or integrity of statistical reports should never be based solely on the STATCHECK algorithm.

**References**

Box, G. E. P. (1954). Some theorems on quadrative forms applied to the study of analysis of variance problems; I: Effect of inequality of variance in the one-way classification. *Annals of Mathematical Statistics, 25,* 484-498.

Geisser, S., & Greenhouse, S. W. (1958). An extension of Box's result on the use of *F* distributions in multivariate analysis. *Annals of Mathematical Statistics, 29,* 885-891.

Huynh, H., & Feldt, L. S. (1970). Conditions under which mean square ratios in repeated measurement designs have exact *F*-distributions. *Journal of the American Statistical Association, 65,* 1582-1589.

Nuijten, M. B., Hartgerink, C. H. J., van Assen, M. A. L. M., Epskamp, S., & Wicherts, J. M. (2016). The prevalence of statistical reporting errors in psychology (1985-2013). *Behavior Research Methods, 48,* 1205-1226. doi:10.3758/s13428-015-0664-2

Schmidt, T., Hauch, V., & Schmidt, F. (2015). Mask-triggered thrust reversal in the negative compatibility effect. *Attention, Perception & Psychophysics, 77:* 2377-2398. doi:10.3758/s13414-015-0923-4

Seydell-Greenwald, A., & Schmidt, T. (2012). Rapid activation of motor responses by illusory contours. *Journal of Experimental Psychology: Human Perception and Performance, 38,* 1168-1182. http://dx.doi.org/10.1037/a0028767

Wicherts, J. M., Bakker, M., & Molenaar, D. (2011). Willingness to share research data is related to the strength of the evidence and the quality of reporting of statistical results. *PLoS One, 6*, e26828.

**Author note**

Correspondence may be sent to the author at thomas.schmidt@sowi.uni-kl.de. I want to thank Michèle Nuijten and Chris Hartgerink for responding so promptly and candidly to my e-mail. I also want to thank Peer 2, who is out on a mission to point out and correct STATCHECK's calculation errors in countless PubPeer postings. Only you know who you are. Thanks to Peer 3 for posting the only content-related comment on our papers (Footnote 2 is my non-anonymous response to your post). Thanks also to Ivan Oransky from Retraction Watch, and of course Filipp Schmidt for checking $p$ values on short notice.

November 16, 2016

**Postscriptum, May 2017**

Nuijten et al.'s (2016) paper has by now received its own discussion forum on PubPeer (https://pubpeer.com/publications/2B9320F0BCF7929F575AA29450599F). In that forum, Michèle Nuijten has addressed the concern that some results of the paper might be biased by $p$-value corrections, as I suppose they are. With regard to the concerns about the "validity check", she announced to conduct a new search through all the 37,000 papers to check whether the prevalence rates she reported are factually correct. She also announced searching for key terms in addition to "Bonferroni" and "Huynh-Feldt". This was in November 2016. Since then, no new postings on the validity problem have appeared in that forum or on any other site that I am aware of. Still, as far as I can tell, Statcheck is less valid as the authors claim, $p$ = .000105.

In an invaluable blog entry from October 2015, Daniël Lakens has reanalyzed the Nuijten et al. (2016) dataset (http://daniellakens.blogspot.de/2015/10/checking-your-stats-and-some-errors-we.html). He reports that about 10 % of the quarter-million tests in the corpus are indeed inaccurate, but that most of the deviations from the correct $p$ value are small. The distribution is asymmetric, and the $p$ value is more likely to be underestimated than to be overestimated. But about one third of the errors are in the other direction and come from $p$ values that are higher than the program expects – these would include the $p$ value corrections described above. Of course, this information is vital to understanding the results, all the more because Nuijten et al. (2016) do not provide any information about the magnitude or direction of the $p$-value deviations. Indeed, Lakens concludes that most of the deviations are due to copy-paste errors, to the reporting of small $p$ values as exactly .000, to the widespread use of "<" instead of "=", and to failures of the program to correctly recognize the test. More seriously, some of the errors result from unspecified use of one-sided tests. In about 10 % of all errors that result in falsely significant results (and about 0.14 % of all tests), decimals are rounded in the wrong direction, so that an actual $p$ value of, say, .059 is reported as .05 – a practice that is sinister enough but still can at most affect the second decimal (in fact, I consider both values equally unimpressive). In my point of view, Lakens' analyses are not consistent with the assertion that "(…) this alarmingly high error rate can have large consequences" (Nuijten et al., 2016, p. 1205). In fact, it is disconcerting that none of Lakens insights into the dataset entered Nuijten et al.'s final paper that was published a year later.

In October 2016, the German Psychological Society criticized the publication of Statcheck results on PubPeer and demanded the deletion of all incorrect entries (https://www.dgps.de/uploads/media/Stellungnahme_DGPs_statcheck_v04_eng_251016.pdf). This has obviously not happened, and the Statcheck comments, even those that are demonstrably wrong, remain uncorrected.

{For this updated version, I decided to drop some additional commentary that seems obsolete now that the authors have provided an additional validity study.}

**Postscriptum, November 2017**

Nuijten, van Assen, Hartgerink, Epskamp, and Wicherts (2017) have published new data on the validity of statcheck. My commentary to that (Schmidt, 2017) can be found here. Based on the data provided by Nuijten et al., I show that Statcheck has poor sensitivity (.52) and poor validity (phi = .54): In 1,120 tests analyzed by the authors, the program scored 29 hits, committed 19 false alarms, and missed an estimated 27 truly inconsistent tests, while 435 tests went unrecognized. If Statcheck flags a test, it is correct in only 60.4 % of all tests. If a test is truly inconsistent, Statcheck flags it in only 51.8 % of cases. Overall, only an estimated 5.00 % of all tests were actually inconsistent, and the program was clearly biased against flagging them.

**Additional Literature:**

Nuijten, M. B., van Assen, M. A. L. M., Hartgerink, C. H. J., Epskamp, S., & Wicherts, J. M. (2017). The validity of the tool "statcheck" in discovering statistical reporting inconsistencies. *PsyArXiv Preprints.* doi:10.17605/OSF.IO/TCXAJ

Schmidt, T. (2017). Statcheck does not work: All the numbers. Reply to Nuijten et al. (2017). *PsyArXiv. November 22. psyarxiv.com/hr6qy.*

T.S., November 23, 2017